\documentclass[9pt,twocolumn,twoside]{pnas-new}
\usepackage{physics}
\templatetype{pnasresearcharticle} 

\title{Many body study of iron(III) bound human serum transferrin}

\author[a,1]{Hovan Lee}
\author[a]{Cedric Weber}
\author[b,1]{Edward B. Linscott}

\affil[a]{Department of Physics, Faculty of Natural \& Mathematical Sciences, King's College London, London, WC2R2LS, UK}
\affil[b]{Theory and Simulation of Materials (THEOS), École Polytechnique Fédérale de Lausanne, 1015 Lausanne, Switzerland}

\leadauthor{Lee} 

\significancestatement{Transferrins are proteins responsible for transporting metal ions in all vertebrates. However, the iron binding properties of transferrins remain poorly understood. Iron, as a transition metal, forms ions with partially-occupied 3\emph{d} subshells. The electrons within the 3\emph{d} orbitals of iron-bound transferrin are therefore highly localized, and interact with one another in a complex manner that cannot be fully characterized by considering each electron separately. In this work, we make use of dynamical mean field theory, a technique that accounts for the strong interactions between these electrons. This is a higher level of theory than has ever been used to study transferrins. We present novel data on the effective spin, multiplet states and optical spectra of iron-bound human serum transferrin.}

\authorcontributions{HL: Investigation, Writing - Original Draft. EBL: Supervision. All authors: Writing - Review \& Editing}
\authordeclaration{The authors declare no competing interests.}
\correspondingauthor{\textsuperscript{1}To whom correspondence should be addressed. E-mail: hovan.lee@kcl.ac.uk, edward.linscott@epfl.ch}

\begin{abstract}
We present the very first density functional theory + dynamical mean field theory calculations of iron-bound human serum transferrin.
Peaks in the optical conductivity at 250, 300 and 450nm were observed, in line with experimental measurements. Spin multiplet analysis suggests that the ground state is a mixed state with high entropy, indicating the importance of strong electronic correlation in this system's chemistry.
\end{abstract}

\dates{This manuscript was compiled on \today}
\doi{\url{www.pnas.org/cgi/doi/10.1073/pnas.XXXXXXXXXX}}

\begin{document}

\maketitle
\thispagestyle{firststyle}
\ifthenelse{\boolean{shortarticle}}{\ifthenelse{\boolean{singlecolumn}}{\abscontentformatted}{\abscontent}}{}

\dropcap{M}etabolism and regulation of iron plays a fundamental role in the homeostasis of the vast majority of living organisms. Vertebrates in particular are reliant on iron, primarily due to the need to synthesize hemoglobin and myoglobin for oxygen transport in blood, and oxygen storage in muscle cells respectively. Furthermore, an imbalance of iron leads to an extensive number of health problems such as anemia (from iron deficiency) and arrhythmia (from iron overload) amongst others \cite{disorder,disorder2,disorder3,disorder4,disorder5,disorber6}. 

Facilitating the iron regulation in vertebrates are the transferrins - a group of metal binding glycoproteins which mediate the transport of iron through blood plasma. These glycoproteins are formed from single poly-peptide chains with molecular weights of $\sim80$kDa, each containing two metal binding sites. 

\begin{figure}
    \centering
    \hspace*{-1.cm} 
    \includegraphics[scale=0.35]{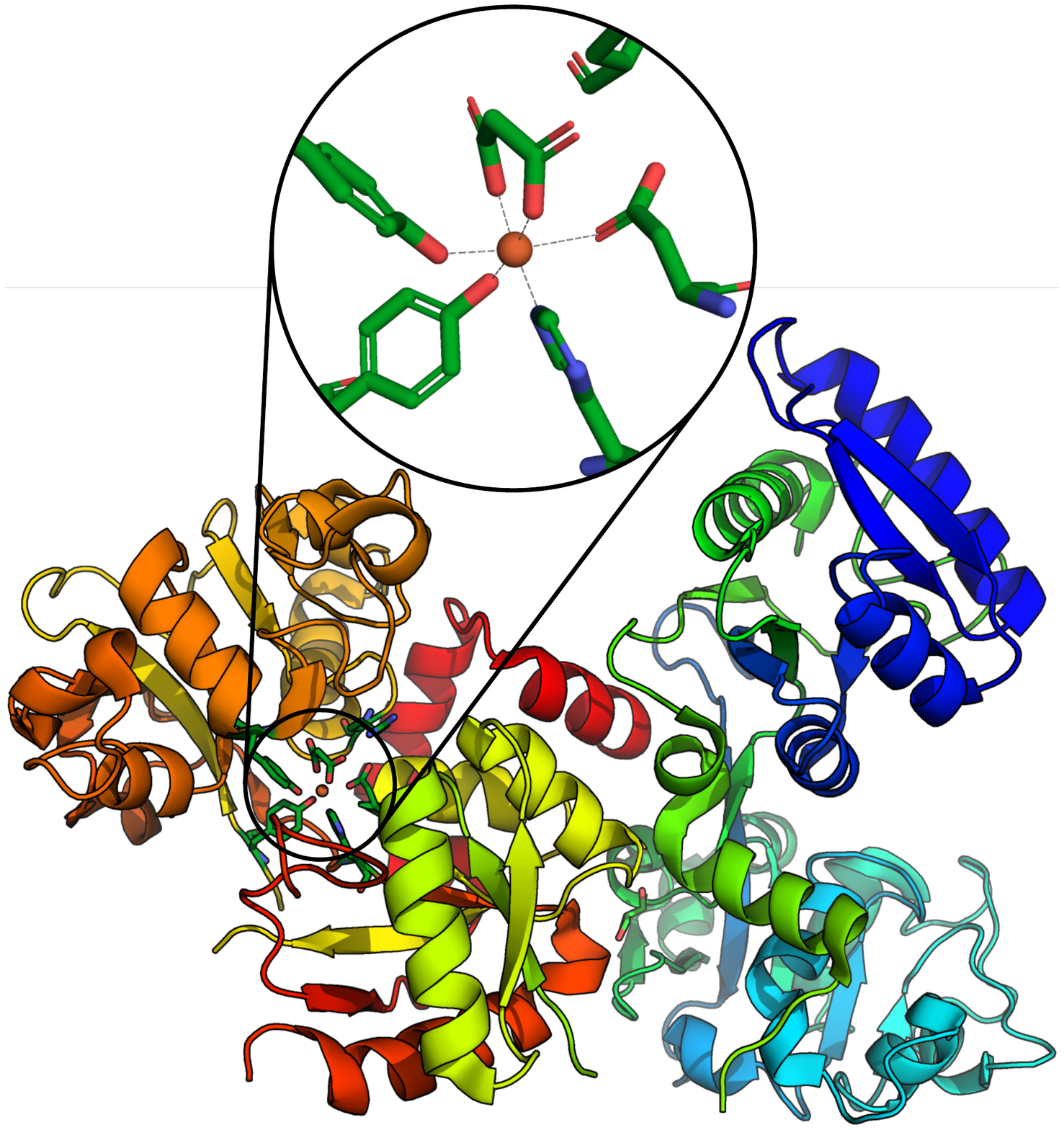}
    \caption{Visualization of the protein sequence of human serum transferrin with a ferric ion bound to the C-lobe \cite{4x1b_structure}. Inset: binding site of the protein, encompassing two tyrosine, one histidine, and one aspartic acid residue, along with the ferric ion and a synergistic malonate anion.}
    \label{fig:toc}
\end{figure}

Crystallographic studies of human serum transferrin (hTF) revealed folded lobes at both the carboxyl- and amino-terminals of the poly-peptide chain (referred to as C-lobe and N-lobe). These lobes contain identical metal binding sites, insofar as the coordination complex amino acids are involved: each site consists of two tyrosine residues, one histidine residue and one aspartic acid residue, with an additional synergistic bidentate anion (such as a carbonate or a malonate ion) to complete the octahedral metal complex, as shown in Fig. \ref{fig:toc}.

Over the past decades, a substantial effort has been made to investigate the metal binding mechanism of hTF. Recent endeavors in X-ray diffraction of crystallized hTF include structures bound with Ti(IV) (both lobes) \cite{ti+4}, Yb(III) (C-lobe) \cite{yb+3} and Cr(III) (C-lobe) \cite{cr+3}. These works have elucidated the conditions and ligand-bond distances needed to accommodate the binding of various ions. 

The conformational changes in hTF have also been studied computationally, using classical molecular dynamics. The latest endeavors include an investigation in the effects of various synergistic and non-synergistic anions on the stability of the binding configuration (Ghanbari et al. \cite{anions}) and an analysis of pH-induced changes on the conformation and its link to the bind/release mechanism of hTF (Kulakova et al. \cite{ph}).

However, classical molecular dynamics do not explicitly model the electrons, as is done in more accurate computational approaches such as density functional theory (DFT). DFT has been used to calculate the binding energies of hTF with various ions: Sanna et al. investigated the different sites responsible for binding VO$^{2+}$ \cite{vo2}, whilst Justino et al. focused on the binding of VO$^{2+}$ in the N-lobe \cite{vo2n}, Sakajiri et al. simulated the binding energies of several different metal ions \cite{sakajiri}, and Reilley et al. examined the uptake and release of a variety of different metal ions \cite{various}.

To calculate the properties of a molecule in DFT, the material is portrayed as an auxiliary system of non-interacting particles. Here the electrons have no explicit influence upon one another, and instead interact at the mean field level, whereby each electron experiences a local potential due to the 
sum of classical Coulombic repulsion between electron densities. 
This level of theory is sufficient in predicting the attributes of a large number of proteins. This is not the case, however, for systems where localized many-body effects are important (such as those containing transition metals) \cite{dft1,dft2,dft3,dft4,dft5,dft6}. Electrons in these systems are in such close proximity to one another that their interactions become too substantial to be treated through the approximation of exchange and correlation in DFT.

One approach to address this issue is to use hybrid functionals. These functionals have an exchange term that is a linear combination of semi-local DFT exchange and Hartree-Fock exchange. The mixing of the two can be carried out such that the problem of self-interaction is minimized. However, hybrid functionals are computationally expensive and scale poorly for large systems such as $\sim 80$kDa proteins. Moreover, hybrid functionals do not include any electronic correlation beyond that contained in the base DFT functional.

In this work, 
we apply dynamical mean field theory (DMFT) \cite{dmft1,dmft2,dmft3,dmft4,dmft5} to study the binding site of hTF. DMFT is a Green's function approach that explicitly calculates the many-body properties of interacting electrons. This approach has found previous success in strongly correlated many-body problems such as explaining the insulating $\mathrm{M}_1$ phase of vanadium dioxide \cite{cedric_vo2}, modeling the photo-disassociation of carboxymyoglobin \cite{oneteptoscam}, and correctly predicting the binding energies of myoglobin \cite{myoglobin} and hemoglobin \cite{hemoglobin} through explicit inclusion of Hund's coupling.


\section*{Results}

\begin{figure}
    \centering
    \includegraphics[width=\linewidth]{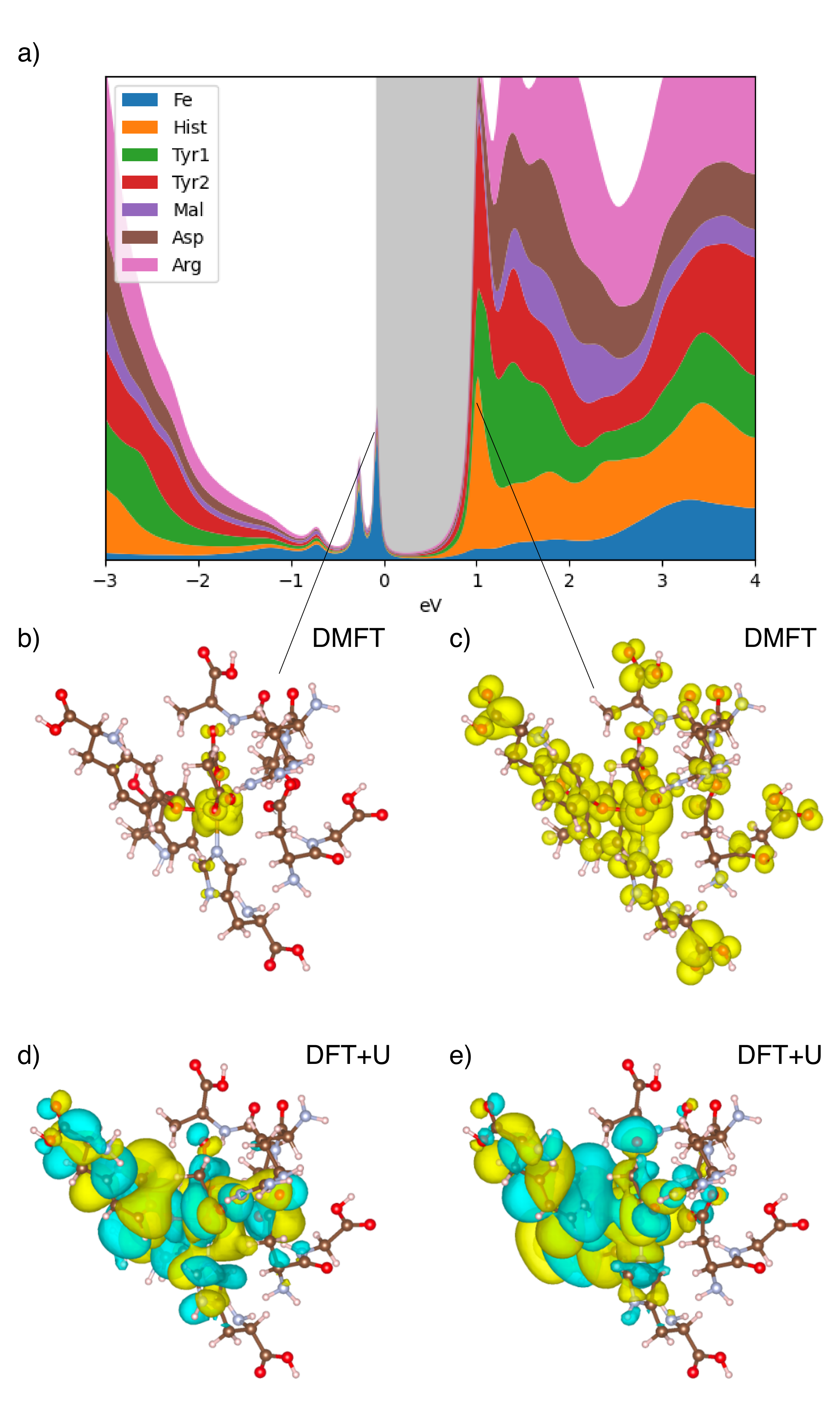}
    \caption{Sub figure a: Density of states (DOS) of the malonate structure, spanning from -3 to +4eV, resolved spatially into the hydrogenic orbitals of the amino acid residues, the Fe ion, and the malonate ion. We observe delocalized, high amplitude features in the DOS spanning from -3 to -1.5eV. Between -1eV and the Fermi level, we observe Fe localized peaks in the LDOS. The lowest unoccupied molecular orbital (LUMO), along with other delocalized higher energy features are seen from $\sim$1eV onward. The real-space resolved DFT+DMFT density of the highest occupied molecular orbital (HOMO) and the LUMO are illustrated in sub figures b and c respectively. The DFT+$U$ spin $=5\hbar/2$ HOMO and LUMO Kohn-Sham orbitals are depicted in sub figures d and e respectively, where yellow (blue) areas indicate positive (negative) parts of the wavefunction.}
    \label{fig:dos}
\end{figure}

DFT+DMFT calculations were performed on three different cluster models of the hTF binding site: one structure with a synergistic malonate anion, and two structures with carbonate anions (referred to in this work as structures CARB A and CARB B).

Firstly, let us examine the spectroscopic properties of hTF as predicted by the DFT+DMFT calculations. The DFT+DMFT local density of states (LDOS) of the malonate structure is shown in Fig. \ref{fig:dos}a. 
The density of states is decomposed into the contributions  of the amino acid residues, the Fe ion, and the malonate ion. Several Fe localized features can be seen below the Fermi level, which is consistent with our expectations: the electrons that are most susceptible to excitations are associated with those of the Fe 3\emph{d} orbital, whilst those of the other atoms (H, C, N, and O) are more tightly bound and are expressed as wide bands below -1eV. Above the Fermi level the lowest unoccupied molecular orbital (LUMO), starting at $\sim$1eV is delocalized throughout the binding site. 

The DFT+DMFT electronic densities of the highest occupied molecular orbital (HOMO) and the LUMO are shown in Fig. \ref{fig:dos}b and c. For comparison, we also show the analogous Kohn-Sham wavefunctions for a DFT+$U$ calculation with broken-spin symmetry and a total spin of $5\hbar/2$. (This state has lower energy than the $\hbar/2$ and $3\hbar/2$ alternatives.)  The DFT+DMFT HOMO is localized around the Fe ion, and the LUMO is delocalized throughout the binding site. This is a qualitatively different picture to what DFT+$U$ predicts, where the HOMO and LUMO are both delocalized throughout the binding site. Clearly, strong local electronic interactions --- as included in DMFT but not in DFT+$U$ --- drive the localization of the HOMO.

\begin{figure}
    \centering
    \includegraphics[width=\linewidth]{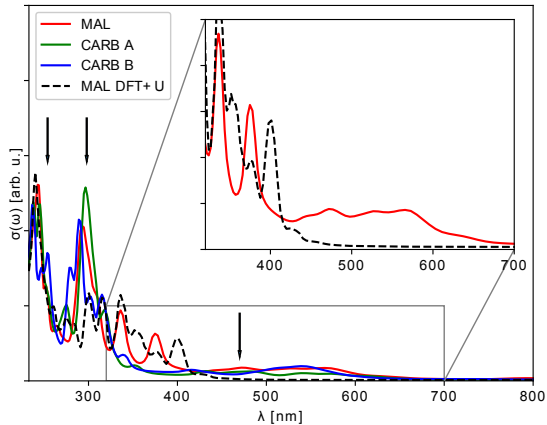}
    \caption{Isotropic optical conductivity, resolved in wavelength, is depicted for the malonate (red), carbonate A (green) and carbonate B (blue) structures. We observe large amplitude features below 250nm and at 300nm, the spectra extend up towards $\sim$700nm. The isotropic optical conductivity for a DFT+$U$ calculation of the malonate structure (dashed black) is also shown. Inset: magnified isotropic optical conductivity spectra. Arrows depict experimentally observed ultraviolet-visible spectroscopic optical absorption bands at 254, 298 and 470nm \cite{uv-vis}.}
    \label{fig:opt}
\end{figure}

Next, we present the optical conductivity spectra of all three hTF structures. These were obtained by applying Kubo-Greenwood relations to the DFT+DMFT density of states. 
The isotropic optical conductivity, defined as $\sigma(\omega)=\frac{1}{3}\sum_{i=x,y,z}\sigma_{ii}(\omega)$, where $\sigma_{ii}(\omega)$ are elements of the frequency dependent optical conductivity matrix, is depicted in Fig. \ref{fig:opt} for the malonate (red), carbonate A (green) and carbonate B (blue) structures. A separate isotropic optical conductivity calculation with DFT+$U$ (dashed black) was also carried out for the malonate structure. 
All three structures concur with the experimentally observed ultraviolet-visible spectroscopy bands: sharp features at 254 and 298nm and a wide band at 470nm \cite{uv-vis}. With DFT+$U$, the feature at $\sim$300nm is not present, and the underlying broadband spectra vanishes at $\sim$450nm. The loss of these UV-VIS features suggests an improper characterization of the correlated 3\emph{d} orbitals of the Fe ion.

\begin{figure}
    \centering
    \includegraphics[width=\linewidth]{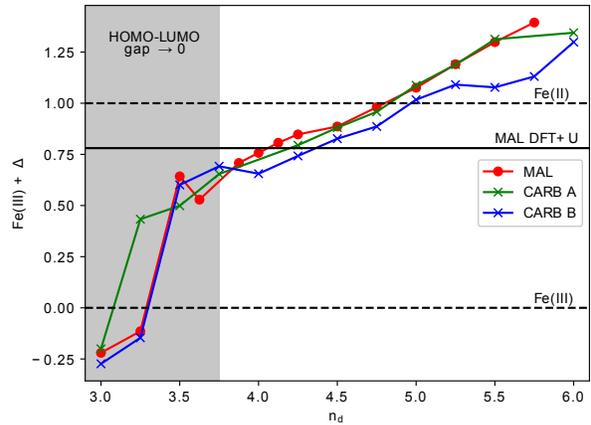}
    \caption{Electronic occupation of the iron 3\emph{d} orbital, shown as $\Delta$: the difference in occupation from Fe(III), for a range of double counting values as applied in the dynamical mean field theory calculations. The reference Fe(III) and Fe(II) electronic occupations are shown as horizontal dashed lines. A solid horizontal line depicts the DFT+$U$ 3\emph{d} orbital filling calculated with a system spin of $5\hbar/2$. For all three structures (malonate in red dots, carbonate form A in green crosses, carbonate form B in blue crosses), the electronic occupation of the orbital increases non-monotonically with double counting values. The resultant range of occupation span from $\sim$Fe(III)$-0.25$ to $\sim$Fe(III)$+1.4$ electrons. The region in grey illustrates DFT+DMFT converged solutions where the gap of the system tends towards zero.}
    \label{fig:dc}
\end{figure}

To summarise, DFT+DMFT predicts spectroscopic properties of hTF that match very well with experiments. However, there is an underlying degree of freedom in the DMFT calculations that we have not yet discussed. This is the double counting parameter \cite{dc}.

The DFT part of these DFT+DMFT calculations treats electronic correlations with delocalized exchange-correlation functionals. The chosen DMFT localized subspace (in this instance, the 3\emph{d} orbitals of iron) are then augmented with screened Coulomb interaction and Hund's coupling explicitly. This calls for a double counting term to correct the component of correlation already included within DFT.

The results of any DFT+DMFT calculation are sensitive to the value of this double-counting parameter. One can in principle determine this parameter from first principles, but we can also treat it as a free parameter that allows us to artificially control the charge of the iron ion. This allows us to explore the electronic state of the iron site in great detail.

We investigate the effects of this double-counting parameter in Fig. \ref{fig:dc}, and report the iron 3\emph{d} orbital occupancy for all three structures in terms of hydrogenic electron counting: Fe(III)$+\Delta$, where Fe(II)=Fe(III)$+1$.
For all three structures, the 3\emph{d} orbital occupation increases with double counting $n_d$, but the two parameters are not directly equivalent: a given double-counting tends to result in a slightly larger 3\emph{d} orbital occupancy. For example, $n_d=3.0$ results in $\sim$Fe(III)$-0.25$, equating to $\sim4.75$ 3\emph{d} electrons. 
In our view, this is due to ligand charge donation: the reshaping of ligand electron wavefunctions upon binding, causing a higher value of electron density around the ion. Nevertheless, the general trend is clear: by changing the double-counting parameter, the occupancy of the iron site -- and more generally, its electronic state -- will change. We also note that the earlier DFT+DMFT results of Fig. \ref{fig:dos} - \ref{fig:opt} used a double-counting parameter that yields the closest 3\emph{d} occupation to the DFT+$U$ converged value of $\sim$Fe(III)$+0.8$.

\begin{figure}
    \centering
    \includegraphics[width=\linewidth]{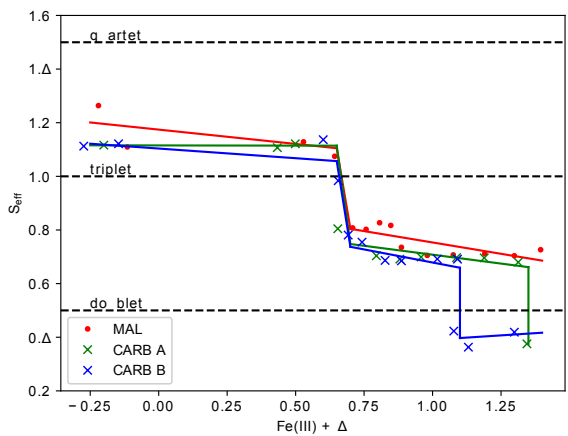}
    \caption{Effective spin value $S_{\mathrm{eff}}$, defined via Tr[$\hat{S^2}\hat{\rho}$] = $\hbar^2S_{\mathrm{eff}}(S_{\mathrm{eff}}+1)$, over the range of 3\emph{d} orbital electronic occupation considered. Horizontal dashed lines are depicted as references to the doublet, triplet and quartet spin states. For all three structures, we observe $S_{\mathrm{eff}}$ values of $\sim1.1$ from $\sim$Fe(III)$-0.25$ to $\sim$Fe(III)$+0.65$. In the case of the malonate structure (red dots), $S_{\mathrm{eff}}$ drops to a value of $0.8$ at Fe(III)$+0.7$ and continues in a declining plateau to a value of $\sim 0.7$ up to Fe(III)$+1.4$. The $S_{\mathrm{eff}}$ trend of the carbonate A structure (green crosses) follows similarly to that of the malonate structure, with an additional data point at $\sim$Fe(III)$+1.3$, $S_{\mathrm{eff}}\sim$0.4, suggesting the onset of a further drop in $S_{\mathrm{eff}}$. $S_{\mathrm{eff}}$ of the carbonate B structure (blue crosses) is similar to that of the other structures, with the addition of a second drop and plateau of $S_{\mathrm{eff}}$ at a value of $\sim$0.4 spanning from $\sim$Fe(III)$+1.1$ to $\sim$Fe(III)$+1.3$. Straight lines are a guide to emphasize the plateaued effective spin values.}
    \label{fig:spin}
\end{figure}

Further exploiting the freedom afforded to us by the double-counting parameter, we explore its effects on the effective spin values of the system in Fig. \ref{fig:spin}.
The effective spin is implicitly defined via the equation $\mathrm{Tr}[\hat{S^2}\hat{\rho}]=\hbar^2S_{\mathrm{eff}}(S_{\mathrm{eff}}+1)$, where $\hat{\rho}$ is the projected 3\emph{d} orbital electronic density matrix and $\hat{S}$ is the dot product between spin vector operators.
In all three structures, we observe plateaus of decreasing $S_\mathrm{eff}$ as 3\emph{d} orbital occupancy increases. 

We noted in Fig. \ref{fig:dc} that the DFT+$U$ calculation converged to a 3\emph{d} occupancy of $\sim$Fe(III)$+0.8$, and that the DFT+DMFT solutions of <Fe(III)$+0.7$ converges to unphysical electronic structures with negligible HOMO-LUMO gap. Therefore, we surmise that the $S_\mathrm{eff}\sim 0.7$ plateau corresponds to the physically meaningful subset of results.



\begin{figure}
    \centering
    \includegraphics[width=\linewidth]{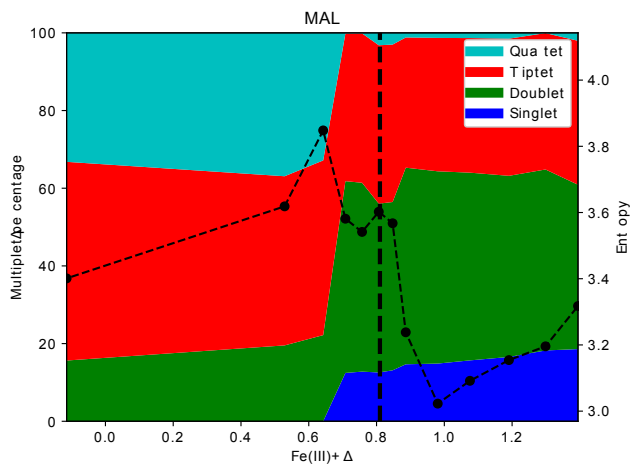}
    \caption{Multiplet analysis of the ground states of the malonate structure calculations, spanning from Fe(III)$-0.1$ to Fe(III)$+1.4$, resolved in the percentages of different contributing multiplet basis states. We remark that the ground states of the calculations from Fe(III)$-0.1$ to Fe(III)$+0.65$ consist of $\sim20\%$ doublet, $\sim40\%$ triplet and $\sim40\%$ quartet states. A sharp spin switch occurs as the electronic occupation increases, and the ground states of the calculations from Fe(III)$+0.7$ to Fe(III)$+1.4$ are contributed primarily by $\sim20\%$ singlet, $\sim40\%$ doublet, $\sim40\%$ triplet and $<5\%$ quartet. Dashed vertical line indicates the value of 3\emph{d} occupancy used in Fig. \ref{fig:dos} and Fig. \ref{fig:opt}. 
    Black dots with dashed line illustrates the entropy associated with the multi-determinant state of the calculations.}
    \label{fig:mult}
\end{figure}


This $S_\mathrm{eff}$ value of $0.7$, and the $S_\mathrm{eff}$ value of other plateaus do not neatly align with the half-integer values of pure multiplet states. Therefore, as a final effort to further characterize the electronic state of the iron in hTF, we expanded the ground state of the malonate structure calculations in terms of their multiplet contributions (Fig. \ref{fig:mult}). We observe that over the range of 3\emph{d} orbital occupancy, 
the ground states encompasses singlet, doublet, triplet and quartet basis states, whilst other multiplet states have negligible contributions. 
The entropy associated with the multi-determinant ground states is also depicted as black dots and a dashed black line. At approximately Fe(III) + 0.7 we see a sharp change in the multiplet contributions, and a simultaneous peak in the entropy. This increased entropy indicates that an increased number of electronic configurations contribute to the electronic state of the 3\emph{d} orbital near this transition. 
Crucially, if we focus on the physically meaningful value of the double-counting (Fe(III) + 0.8, indicated with a vertical dashed line), we can conclude that the spin state comprises of singlet, doublet, triplet, and quartet contributions, and that the iron resides in a region of high entropy. This is a signature of strong electronic correlation.

Before closing, we should note that experimental measurements of iron-bound hTF, including M\"{o}ssbauer \cite{moss+epr,moss1} and electron paramagnetic resonance (EPR) spectroscopy \cite{moss+epr,epr1,epr2}, produces results that indicate a high-spin Fe(III) sextet: $S_\mathrm{eff}=5/2$. From our perspective, the main contributing factor of the difference between our $S_\mathrm{eff}$ values and these experiments stem from the experimental requirement of low temperatures and high external magnetic fields; as an example, the EPR experiment of \cite{epr2} was performed under external fields spanning $7.5$ to $12\mathrm{T}$, 
up to a temperature of $90\mathrm{K}$. 
In comparison, our calculations were performed at $293\mathrm{K}$ in the absence of external magnetic fields, allowing us to identify the superposition of different multiplet states. This superposition of states would inevitably be disassembled as temperature decreases and external field strength increases (due to alignment of spins to magnetic field), exciting the system out of its mixed state as it pertains to in vivo conditions, resulting in the experimentally observed $S_\mathrm{eff}=5/2$.

\section*{Conclusion}

In this work, we utilized DFT+DMFT to calculate the electronic properties of the transitional metal binding site of human serum transferrin. We investigated three cluster models, all three structures exhibit a tendency of increasing the electronic occupation of Fe(III) when it becomes bound to the binding site. The DFT+DMFT local density of states revealed a HOMO-LUMO gap of $\sim$ 1eV, with the HOMO states localized at the Fe ion. The optical conductivity of the systems were calculated, with discernible features at 250, 300 and 450nm, coinciding with the experimentally observed absorption peaks at 254, 298 and 470nm.
Further inspections into the systems' multiplet contributions revealed that the ground states consist of singlet, double, triplet and quartet states. Moreover, the system is in a region of high entropy. This indicates that strong electronic correlation plays an important role in the electronic state of iron in hTF.

\subsection*{Methods}

Crystallographic structures of carbonate (N-lobe) and malonate (C-lobe) Fe(III)-hTF were obtained from the Protein Data Bank archives \cite{1a8e_structure} and \cite{4x1b_structure} respectively. In the case of carbonate hTF, the crystallized orthorhombic ($P2_12_12_1$) structure contains alternative conformations of the synergistic carbonate anion, these structures are referred to in this work as structures A and B.

The binding sites of these structures were inspected, and cluster models were defined with a radius of 33{\AA}, centered at the Fe ion. The clusters were then geometrically optimized, with the Fe ion, ligand donor atoms, and the amino acid backbone fixed.

To perform this geometry optimisation we used the linear-scaling DFT software ONETEP \cite{onetep1,onetep2}.  In ONETEP, DFT calculations are based on minimizing the system energy with respect to the single particle density matrix:
\begin{equation}
    \rho(\textbf{r},\textbf{r'})=\sum_{\alpha,\beta}\phi_\alpha(\textbf{r})K^{\alpha,\beta}\phi_\beta(\textbf{r'})
\end{equation}
where $\{\phi_\alpha\}$ refers to a set of non-orthogonal generalized Wannier functions (NGWFs), and $K^{\alpha\beta}$ is the density kernel.

All DFT calculations were performed using the PBE XC functional \cite{Perdew1996a}, with an energy cutoff at 900eV. There were thirteen NGWFs on the iron atom, four on each carbon, nitrogen, and oxygen, and one on each hydrogen. All NGWFs had a 6.5{\AA} cutoff radius. A padded cell \cite{Hine2011b} of 34{\AA}  was applied to produce open boundary conditions. 
The systems were embedded in an implicit solvent with relative bulk permittivity of 80 \cite{is1}. This approach avoids the spurious closing of the HOMO-LUMO gap that can occur in cluster models of proteins \cite{is2}.

Paramagnetic DFT+DMFT calculations were then performed on the optimized structure. This involves self-consistently calculating the Green's function as follows.

To start, a converged DFT calculation yields the NGWF basis Kohn-Sham Hamiltonian $H^{\alpha\beta}$, from which the NGWF resolved Green's function is obtained:
\begin{equation}
    G^{\alpha\beta}(\omega)=[(\omega+\mu)S-H-\Sigma(\omega)]_{\alpha\beta}^{-1}
\end{equation}
where $\omega$ can be cast into $\omega+i\eta$ or $i\omega_n$ for broadening or finite-temperature Matsubara calculations, $\mu$ is the chemical potential, $S_{\alpha\beta}$ is the NGWF overlap matrix, and $\Sigma_{\alpha\beta}$ is the self energy.

In order to reintroduce the physics of strong correlation that is absent in DFT, the system is then downfolded onto an Anderson impurity model (AIM):
\begin{multline}
\hat H =
\underbrace{\sum_{ij\sigma} (\varepsilon_{ij}-\mu) \hat c^\dag_{i\sigma} \hat c_{j\sigma}}_{\hat H_\text{bath}} 
+ \underbrace{\sum_{im\sigma} \left(V_{m i}\hat f^\dag_{m\sigma} \hat c_{i\sigma}
+ h.c.\right)}_{\hat H_\text{mix}} \\
+ \underbrace{
\sum_{mm'\sigma} (t_{mm'}-\mu)\hat f^\dag_{m\sigma}\hat f_{m'\sigma} + \hat H_{U}
}_{\hat H_\text{loc}}
\end{multline}
where $\hat{H}_{bath}$ describes the non-correlated parts of the system (with hopping parameter $\epsilon_{ij}$), $\hat{H}_{loc}$ characterizes the correlated orbitals by introducing the physics of strong correlation to the iron 3\emph{d} subspace (with hopping parameter $t_{mm'}$ and Slater-Kanamori Hamiltonian $\hat{H}_U$) and $\hat{H}_{mix}$ represents the coupling in-between. From this, the non-interacting impurity Green's function is defined as:
\begin{equation}
    {{G}_{imp_{mm'}}^0}(\omega)^{-1}
    = \omega +\mu - t - 
    V_{mi} \left(\frac{1}{\omega+\mu-{\varepsilon}}\right)_{ij} V^{\dag}_{jm'}
\end{equation}
and the Green's function of the AIM
\begin{multline}
        G_{imp_{mm'}}(\omega)=\left\langle \hat f_m \frac{1}{\omega -(\hat H - E_0)} \hat f^\dag_{m'} \right\rangle \\
  + \left\langle \hat f^\dag_{m'} \frac{1}{\omega +(\hat H - E_0)} \hat f_m \right\rangle 
\end{multline}
was calculated through the use of an exact diagonalization Lanczos solver \cite{lanczos}.

The impurity self-energy is then obtained via Dyson's equation:
\begin{equation}
    \Sigma_\text{imp}(\omega)=[G^0_\text{imp}(\omega)]^{-1}-G_\text{imp}^{-1}(\omega)
\end{equation}
This self-energy is then upfolded back into the NGWF basis, whereby properties such as the local density of states and the optical conductivity can be extracted.

The original DFT calculation already treats Coulomb interaction to some degree, we therefore need to apply corrections such that the Coulomb interactions are not double counted. $n_d$, as it is shown in Fig. \ref{fig:dc} is accounted for as:
\begin{equation}
    E_{DC}=\frac{U+2(N-1)(U-2J)}{2N-1}(n_d-0.5)+\frac{J}{2}(n_d-1)
\end{equation}
where N is the number of orbitals spanning the correlated subspace, $U$ is the Hubbard $U$ and $J$ the Hund's exchange parameter. $E_{DC}$ is the double counting correction to the impurity self-energy: $\Sigma_{imp,DC}=\Sigma_{imp}-E_{DC}$

The paramagnetic DFT+DMFT calculations presented in this work were performed using ONETEP and the DMFT package TOSCAM \cite{toscam}. The calculations used a Slater-Kanamori Hubbard $U$ and Hund's exchange $J$ of 4.0eV and 0.8eV respectively. Seven bath sites, along with the CPT extension \cite{cpt} were used in all DFT+DMFT calculations. The DFT+$U$ results presented in Fig. \ref{fig:dos} - \ref{fig:dc} were calculated with a matching Hubbard parameter $U_\mathrm{eff}$ and exchange parameter $J$ values of 4.0eV and 0.8eV respectively \cite{dft+u,dft+u2,dft+u3}. These calculations were spin-polarized with a total magnetization of $5\hbar/2$, corresponding to the lowest-energy spin configuration.

The iron 3\emph{d} orbital occupation, as presented in Fig. \ref{fig:dc} - \ref{fig:mult} is obtained via projecting the DMFT converged ground state onto the Kohn-Sham solutions for a lone iron scalar relativistic pseudopotential generated with OPIUM \cite{opium,Kerker1980a,Kleinman1982a,Hamann1989a,Rappe1990a,Gonze1991a,Ramer1999a,Grinberg2000a}.

The effective spin (as presented in Fig. \ref{fig:spin}) is calculated via the reduced density matrix $\hat \rho = \sum_i e^{-\beta E_i} \mathrm{Tr}_B[\ket{i}\bra{i}]$, whereby a partial trace was taken over the bath degrees of freedom in the AIM. We can then define the effective spin of the iron atom, $S_{\mathrm{eff}}$, via the expectation value of $\hat S^2 = \sum_{i,j} \hat{\mathbf{S}}_i \cdot \hat{\mathbf{S}}_j$ and the relation $\mathrm{Tr}[\hat S^2 \hat \rho] = \hbar^2 S_\mathrm{eff}(S_\mathrm{eff} + 1$).

The von Neumann entropy (as presented in Fig. \ref{fig:mult}) is given by:
\begin{equation}
    S(\hat{\rho})=-Tr[\hat{\rho}\mathrm{ln}(\hat{\rho})]
\end{equation}

Multiplet contributions to the DMFT converged reduced density matrix (as presented in Fig. \ref{fig:mult}) are obtained from constructing the spin projector:
\begin{equation}
\hat P_S = \sum_ {s \in S} \ket{s} \bra{s}
\end{equation}
where the eigenstates $\ket{s}$, when applied onto the spin operator $\hat{S}^2$, are associated with the eigenvalues $S(S+1)$. The fractions of the reduced density matrix, corresponding to different multiplet states, are obtained via $Tr[\hat{P}_S\hat{\rho}\hat{P}_S]$ for $S$ = $0$,$\frac{1}{2}$,$1$ etc.

Local density of states (as presented in Fig. \ref{fig:dos}) is obtained from upfolding the converged DMFT impurity Hamiltonian into the NGWF basis. 
The NGWF basis resolved DFT+DMFT density matrix is then:
\begin{equation}
    \rho^{\alpha\beta}=\frac{1}{2\pi i}(G^{\alpha\beta}(\omega)-G^{\alpha\beta\dagger}(\omega))
\end{equation}
and the trace of this density matrix can be taken to obtain the LDOS:
\begin{equation}
    \rho_I(\omega)=\sum_{\alpha\in I}\sum_\beta\rho^{\alpha\beta}(\omega)S_{\alpha\beta}
\end{equation}
where $I$ denotes the subset of NGWFs that belong to each ion or amino acid residue.

The optical absorption spectrum of the system (as presented in Fig. \ref{fig:opt}) can be obtained with the linear response Kubo formalism:
\begin{multline}
    \sigma_{ij}(\omega) = \frac{2\pi}{\Omega} \int d\omega' \frac{f(\omega' - \omega) - f(\omega')}{\omega} \\
\times \left(\rho^{\alpha \beta}(\omega' - \omega)\mathbf{v}^i_{\beta \gamma} \rho^{\gamma \delta}(\omega') \mathbf{v}^j_{\delta \alpha}\right)
\end{multline}
where $\Omega$ is the simulation cell volume, $f(\omega)$ is the Fermi-Dirac distribution, $\rho$ is the NGWF basis spectral density, $i$ and $j$ indices indicate Cartesian directions, the velocity operator $\mathbf{v}$ is defined as:
\begin{equation}
\mathbf{v}^j_{\alpha \beta} = -i\bra{\alpha} \nabla_j \ket{\beta} + i \bra{\alpha} \left[\hat V_{nl}, \mathbf{r}\right]\ket{\beta}
\end{equation}
with non-local pseudopotentials $\hat V_{nl}$.

\acknow{CW was supported by grant EP/R02992X/1 from the UK Engineering and Physical Sciences Research Council (EPSRC). EBL gratefully acknowledges financial support from the Swiss National Science Foundation (SNSF -- project number 200021-143636). The authors thank E. Picard, Y. Courtois and F. Behar-Cohen for useful discussions.}

\showacknow{} 

\bibliography{citations}

\end{document}